\documentclass[a4paper,12pt]{article}
\pdfoutput=1 

\usepackage{jcappub} 
\usepackage[symbol]{footmisc}
\usepackage{changes}
\usepackage{amsmath,amssymb}
\usepackage{graphicx}
\usepackage{adjustbox}
\usepackage{multirow}
\usepackage[normalem]{ulem}
\usepackage{bm}        
\newcommand{\fpbh}{f_{\mathrm{PBH}}}

\usepackage{graphicx}   
\usepackage{tabulary}
\usepackage{color}

\newcommand{\nc}{\newcommand*}

\nc{\al}{\alpha}
\nc{\s}{\sigma}
\nc{\dt}{\delta}
\nc{\Dt}{\Delta}
\nc{\Ld}{\Lambda}
\nc{\p}{\partial}
\nc{\om}{\omega}
\nc{\Om}{\Omega}
\nc{\rd}{\mathrm{d}}
\nc{\Od}[1]{\mathcal{O}(#1)} 
\nc{\kp}{\kappa}

\def\({\left(}
\def\){\right)}
\def\[{\left[}
\def\]{\right]}
\def\e{\begin{equation}}
\def\q{\end{equation}}
\def\m{\begin{eqnarray}}
\def\n{\end{eqnarray}}

\nc{\Eq}[1]{Eq.~\eqref{#1}}     
\nc{\Fig}[1]{Fig.~\ref{#1}}     
\nc{\Table}[1]{Table~\ref{#1}}  
\nc{\Sec}[1]{Sec.~\ref{#1}}     

\nc{\Msun}{M_\odot}             
\nc{\fpbhn}{f_{\mathrm{pbh0}}}    
\nc{\mR}{\mathcal{R}} 
\nc{\seq}{\sigma_{\mathrm{eq}}}
\nc{\ogw}{\Omega_{\mathrm{GW}}}
\nc{\gpcyr}{\mathrm{Gpc}^{-3}\,\mathrm{yr}^{-1}}
\nc{\lvc}{LIGO/Virgo} 
\nc{\SNR}{\mathrm{SNR}} 
\nc{\mmin}{{m_{\mathrm{min}}}}
\nc{\mmax}{{m_{\mathrm{max}}}}
\nc{\Mmin}{{M_{\mathrm{min}}}}
\nc{\fmin}{{f_{\mathrm{min}}}}
\nc{\VT}{\mathrm{VT}}
\nc{\rhoGW}{\rho_{\mathrm{GW}}}
\nc{\vth}{\vec{\theta}}
\nc{\vd}{\vec{d}}
\nc{\vla}{\vec{\lambda}}
\nc{\Nobs}{N_{\mathrm{obs}}}
\nc{\av}[1]{\langle #1 \rangle} 
\nc{\km}{\mathrm{km}}
\nc{\Mpc}{\mathrm{Mpc}}
\nc{\Tobs}{T_{\mathrm{obs}}}
\nc{\Ntemp}{N_{\mathrm{temp}}}
\nc{\ie}{\textit{i.e.}}
\nc{\addref}{[\textcolor{red}{add ref}] } 
\nc{\eg}{\textit{e.g.~}}
\nc{\app}{\approx}
\nc{\hf}{\frac{1}{2}}
\nc{\discuss}{\textcolor{red}{Add discussion here!}}

\nc{\mpbh}{m_{\rm{pbh}}}
\nc{\cR}{\mathcal{R}}
\nc{\mU}{{\mathcal{U}}}
\nc{\Mc}{{M_\mathrm{c}}}
\nc{\Mf}{{M_\mathrm{f}}}
\nc{\red}[1]{\textcolor{red}{#1}}
\nc{\yellow}[1]{\textcolor{yellow}{#1}}
\nc{\green}[1]{\textcolor{green}{#1}}
\nc{\blue}[1]{\textcolor{blue}{#1}}
\nc{\fnl}{F_{\mathrm{NL}}}
\nc{\gnl}{G_{\mathrm{NL}}}
\nc{\MG}{\mathcal{M}_{\mathrm{G}}}
\nc{\MNG}{\mathcal{M}_{\mathrm{NG}}}

\begin{document}
	
\title{Probing the equation of state of the early Universe with pulsar timing arrays} 

\author{Lang~Liu,$^{a,b}$}
\author{Zu-Cheng~Chen,\note{Corresponding author.}$^{a,b,c,*}$}
\author{and Qing-Guo Huang$^{d,e,f,*}$}

\affiliation{$^a$Department of Astronomy, Beijing Normal University, Beijing 100875, China}
\affiliation{$^b$Advanced Institute of Natural Sciences, Beijing Normal University, Zhuhai 519087, China}
\affiliation{$^c$Department of Physics and Synergistic Innovation Center for Quantum Effects and Applications, Hunan Normal University, Changsha, Hunan 410081, China}
\affiliation{$^d$CAS Key Laboratory of Theoretical Physics, Institute of Theoretical Physics, Chinese Academy of Sciences, Beijing 100190, China}
\affiliation{$^e$School of Physical Sciences, University of Chinese Academy of Sciences, No. 19A Yuquan Road, Beijing 100049, China}
\affiliation{$^f$School of Fundamental Physics and Mathematical Sciences, Hangzhou Institute for Advanced Study, UCAS, Hangzhou 310024, China}

\emailAdd{liulang@bnu.edu.cn}	
\emailAdd{zucheng.chen@bnu.edu.cn}
\emailAdd{huangqg@itp.ac.cn}
	
\abstract{
The recently released data by pulsar timing array (PTA) collaborations have amassed substantial evidence corroborating the existence of a stochastic signal consistent with a gravitational-wave background at frequencies around the nanohertz regime. We investigate the situation in which the PTA signal originates from scalar-induced gravitational waves~(SIGWs), which serves as a valuable tool to probe the equation of state parameter $w$ during the Universe's early stages. The joint consideration of the PTA data from the NANOGrav 15-year data set, PPTA DR3, and EPTA DR2 yields that $w=0.60^{+0.32}_{-0.39}$ (median + $90\%$ credible interval), indicating a period of condensate domination at the production of SIGWs {is allowed by the data. Moreover, the data also supports radiation domination ($w=1/3$) within the $90\%$ credible interval}. We also impose an upper bound on the reheating temperature that $T_\mathrm{rh} \lesssim 0.2\,\mathrm{GeV}$ and the constraint on $w$ reveals valuable information on the inflationary potential and the dynamics at the end of inflation. 
}
	
\maketitle
\section{Introduction}

Recently, four major pulsar timing array (PTA) collaborations, namely the North American Nanohertz Observatory for Gravitational Waves (NANOGrav)~\cite{NANOGrav:2023gor,NANOGrav:2023hde}, the Parkes PTA (PPTA)~\cite{Reardon:2023gzh,Zic:2023gta}, the European PTA (EPTA) along with the Indian PTA (InPTA)~\cite{Antoniadis:2023rey,Antoniadis:2023utw}, and the Chinese PTA (CPTA)~\cite{Xu:2023wog}, have jointly announced compelling evidence of a common-spectrum signal consistent with the Hellings-Downs spatial correlations~\cite{Hellings:1983fr} in their latest data sets, thus suggesting a gravitational wave (GW) origin for this signal. While various physical phenomena~\cite{Li:2019vlb,Chen:2021wdo,Wu:2021kmd,Chen:2021ncc,Chen:2022azo,Liu:2022lvz,PPTA:2022eul,Wu:2023pbt,Madge:2023cak,Wu:2023dnp,Liu:2023hte,IPTA:2023ero,Wu:2023rib,Bi:2023ewq,Chen:2023uiz} can produce GW signals within the PTA frequency band, the precise origin of this signal, whether attributed to supermassive black hole binaries~\cite{NANOGrav:2023hfp, Antoniadis:2023xlr,Bi:2023tib,Cannizzaro:2023mgc} or other cosmological sources~\cite{NANOGrav:2023hvm,Datta:2023vbs,Vagnozzi:2023lwo,Han:2023olf,Li:2023yaj,Franciolini:2023wjm,Shen:2023pan,Kitajima:2023cek,Franciolini:2023pbf,Addazi:2023jvg,Cai:2023dls,Inomata:2023zup,Murai:2023gkv,Li:2023bxy,Anchordoqui:2023tln,Liu:2023ymk,Abe:2023yrw,Ghosh:2023aum,Figueroa:2023zhu,Yi:2023mbm,Wu:2023hsa,Li:2023tdx,Geller:2023shn,You:2023rmn,Antusch:2023zjk,Ye:2023xyr,HosseiniMansoori:2023mqh,Jin:2023wri,Zhang:2023nrs,ValbusaDallArmi:2023nqn,DeLuca:2023tun,Gorji:2023sil,Das:2023nmm,Yi:2023npi,Yi:2023tdk,InternationalPulsarTimingArray:2023mzf,Chen:2023zkb}, remains an ongoing subject of investigation.

A plausible explanation for the observed signal is the presence of scalar-induced gravitational waves (SIGWs), which are generated by the primordial curvature perturbations at small scales~\cite{Ananda:2006af,Baumann:2007zm,Garcia-Bellido:2016dkw,Inomata:2016rbd,Garcia-Bellido:2017aan,Kohri:2018awv,Cai:2018dig,Lu:2019sti,Yuan:2019wwo,Chen:2019xse,Xu:2019bdp,Cai:2019cdl,Yuan:2019fwv,Yi:2020kmq,Yi:2020cut,Liu:2020oqe,Gao:2020tsa,Sakharov:2021dim,Yi:2021lxc,Yi:2022anu,Yi:2022ymw,Meng:2022low}. As these curvature perturbations attain significant magnitudes, they can give rise to a substantial stochastic gravitational-wave background (SGWB) through second-order effects arising from the non-linear coupling of perturbations. Moreover, large curvature perturbations can trigger the formation of primordial black holes (PBHs)~\cite{Zeldovich:1967lct, Hawking:1971ei,Carr:1974nx}. PBHs have attracted considerable attention in recent years~\cite{Saito:2008jc,Belotsky:2014kca,Carr:2016drx,Garcia-Bellido:2017mdw,Carr:2017jsz,Germani:2017bcs,Chen:2018rzo,Liu:2018ess,Chen:2018czv,Liu:2019rnx,Fu:2019ttf,Cai:2019elf,Liu:2019lul,Cai:2019bmk,Yuan:2019udt,Chen:2019irf,Liu:2020cds,Wu:2020drm,Fu:2020lob,DeLuca:2020sae,Vaskonen:2020lbd,DeLuca:2020agl,Domenech:2020ers,Hutsi:2020sol,Chen:2021nxo,Liu:2021svg,Kawai:2021edk,Braglia:2021wwa,Liu:2021jnw,Braglia:2022icu,Chen:2022qvg,Zheng:2022wqo,Liu:2022iuf,Chen:2022fda,Gu:2022pbo,Guo:2023hyp,Gu:2023mmd} because they can not only be a compelling candidate for dark matter~\cite{Sasaki:2018dmp,Carr:2020gox,Carr:2020xqk} but also explain the binary black holes detected by LIGO-Virgo-KAGRA~\cite{Bird:2016dcv,Sasaki:2016jop}.

The infrared power-law index of SIGWs is susceptible to the equation of state (EoS) of the early Universe when the corresponding wavelength reenters the Hubble horizon~\cite{Domenech:2019quo,Domenech:2020kqm,Domenech:2021ztg}. While the standard cosmological model posits that the Universe was dominated by radiation from the end of inflation until the onset of matter domination, the only well-established evidence comes from the big bang nucleosynthesis (BBN) period, during which the Universe was radiation-dominated. However, prior to BBN, there was limited observational data, providing an opportunity to explore phenomena associated with a different EoS parameter ($w\neq1/3$) during that earlier stage~\cite{Allahverdi:2020bys}. This offers a promising pathway for investigating various scenarios that could have occurred before the BBN era. Therefore, SIGWs offer an invaluable opportunity to  directly investigate the expansion history of the primordial dark Universe.
There is considerable room for new physics to manifest, allowing intriguing phenomena to emerge in the context of the SIGWs. Various aspects of the SIGWs provide avenues for potential new physics to be explored, such as investigating different equations of state for the Universe~\cite{Domenech:2020ssp,Domenech:2021wkk,Domenech:2019quo,Domenech:2020kqm}, exploring different propagation speeds of fluctuations~\cite{Domenech:2021ztg,Balaji:2022dbi,Balaji:2023ehk}, and investigating different initial conditions~\cite{Domenech:2021and}. These fascinating scenarios can be tested and probed using the SIGWs as a powerful observational tool.

In this paper, we explore a scenario where the Universe, {prior to the end of reheating}, is characterized by an arbitrary constant EoS parameter $w$ and a propagation speed $c_s^2=1$. Such conditions can be realized by a scalar field either oscillating in an arbitrary potential or rolling down an exponential potential~\cite{Lucchin:1984yf}. In the context of the SIGWs, the infrared power-law index experiences a change at the reheating frequency, corresponding to the mode that reenters the horizon when the Universe transitions into a radiation-dominated phase. Assuming that the stochastic signal detected by PTAs emanates from SIGWs, we conduct a comprehensive analysis by employing PTA data jointly from the NANOGrav 15-year data set, PPTA DR3 and EPTA DR2. Our primary objective is to constrain the EoS parameter of the early Universe as well as the reheating energy scale through PTA data.
	
\section{Scalar-induced gravitational waves and primordial black holes} 

\begin{figure}[tbp!]
	\centering
	\includegraphics[width=\textwidth]{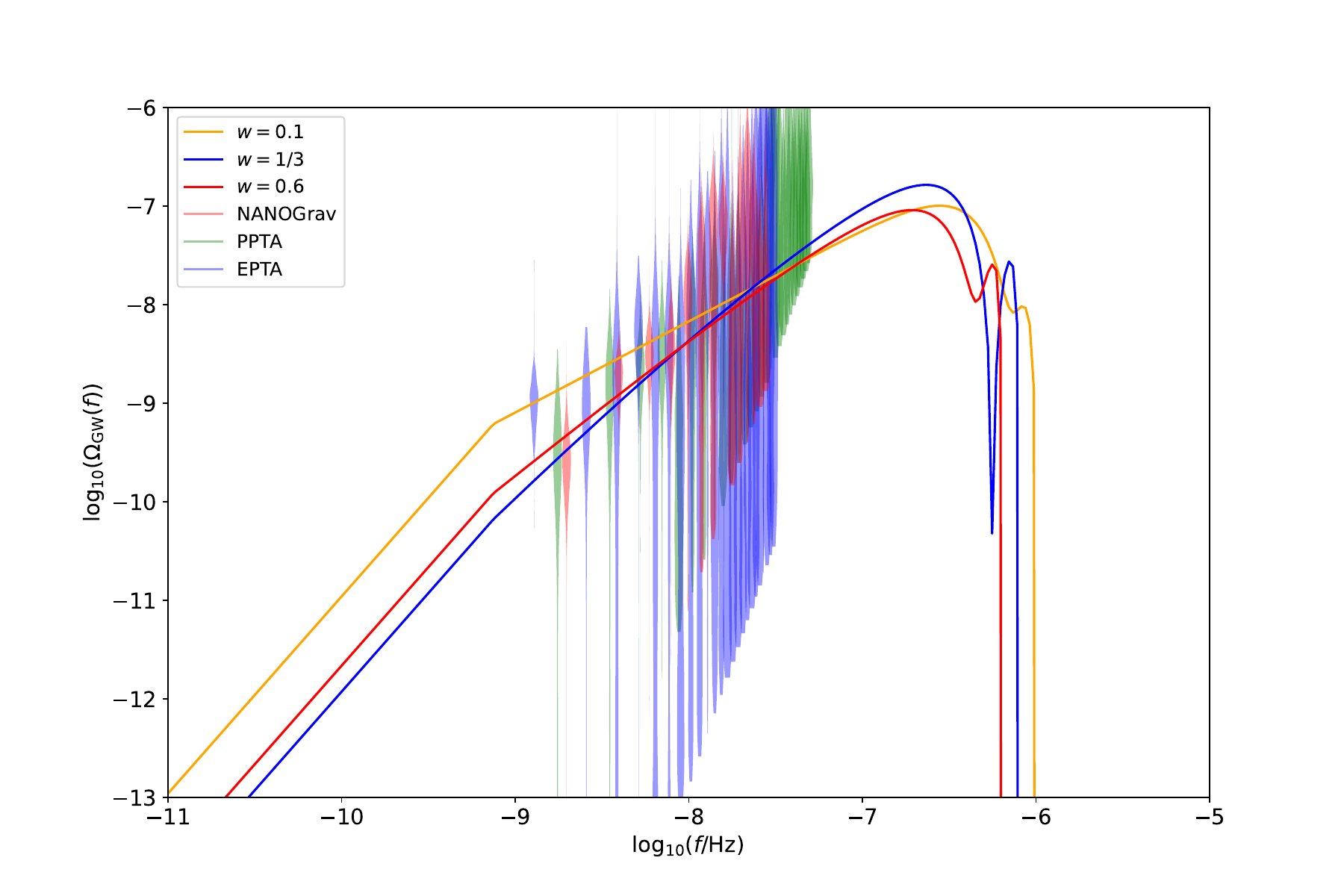}
	\caption{\label{ogw} The energy density spectrum of SIGW with different EoS parameters as a function of GW frequency.}
\end{figure}

The scalar perturbations must exhibit a substantial enhancement compared to the fluctuations observed in the cosmological microwave background (CMB) experiments to generate a detectable SGWB. We adopt a log-normal form for the primordial power spectrum of curvature perturbations $\mathcal{P}_{\mathcal{R}}$ that is commonly used in the literature~\cite{Kohri:2018awv, Ferrante:2022mui}
\begin{equation}
\label{PR}
\mathcal{P}_{\mathcal{R}} (k) = \frac{A}{\sqrt{2\pi}\Delta} \exp \left( -\frac{\ln^{2}(k/k_{*})}{2\Delta^{2}} \right),
\end{equation}
where $A$ is the amplitude, $k_*$ is the characteristic scale, and $\Delta$ is width of the spectrum.
In the following, we restrict our attention to the narrow peak spectrum, namely, $\Delta \lesssim0.1$, which is supported by our results. In the limit as $\Delta\rightarrow 0$, the primordial power spectrum $\mathcal{P}_{\mathcal{R}}$ takes the form of a $\delta$-function,  $\mathcal{P}_{\mathcal{R}}=A k_*\delta(k-k_*)$. The effect of a finite width for SIGW can be expressed as~\cite{Pi:2020otn}
\begin{equation}\label{eq:OmegaDelta}
\Omega_{\rm GW,0}h^2={\rm Erf}\left[\frac{1}{\Delta}\sinh^{-1}\frac{k}{2k_{\rm *}}\right]\Omega^\delta_{\rm GW,0}h^2.
\end{equation}
where $\Omega^\delta_{\rm GW,0}$ is the GW spectrum induced by a $\delta$-function peak. To calculate the waveform of SIGW in a $w$-dominated universe, we rely on the analytical results presented in Ref.~\cite{Domenech:2020kqm,Domenech:2021ztg}. Additionally, we consider an instantaneous reheating scenario, where the Universe reheats immediately when the $k_\mathrm{rh}$-mode reenters the horizon. The observed spectrum of SIGW per $\ln k$ today is
\begin{equation}\label{ogw0}
\Omega_{\mathrm{GW}, 0} h^2 \approx  1.62 \times 10^{-5} \left(\frac{g_{*r}\left(T_{\mathrm{rh}}\right)}{106.75}\right)\left(\frac{g_{* s}\left(T_{\mathrm{rh}}\right)}{106.75}\right)^{-\frac{4}{3}} \left(\frac{\Omega_{r, 0} h^2}{4.18 \times 10^{-5}}\right)  \Omega_{\mathrm{GW}, \mathrm{rh}}, 
\end{equation}
where $\Omega_{r,0}h^2 \approx 4.18 \times 10^{-5} $ is the radiation fraction today,  $g_{*r}$ and $g_{*s}$ are the effective energy and entropy degrees of freedom, respectively. The SIGW  spectrum for the scales $k \gtrsim k_{\rm rh}$ in Eq.~\eqref{ogw0} is
\begin{equation}\label{ogwrh}
\Omega_{\mathrm{GW},\mathrm{rh}}=\left(\frac{k}{k_{\mathrm{rh}}}\right)^{\!\!\! -2 b}\!\!\!\! \int_0^{\infty}\!\! d v\!\! \int_{|1-v|}^{1+v}\!\! d u\, \mathcal{T}\, \mathcal{P}_{\mathcal{R}}(k u) \mathcal{P}_{\mathcal{R}}(k v),
\end{equation}
where $b\equiv (1-3w)/(1+3w)$, and the detailed expression for $\mathcal{T}\equiv\mathcal{T}(u, v, w)$ can be found in~\cite{Domenech:2020kqm,Domenech:2021ztg}. We have $\Omega_{\mathrm{GW},\mathrm{rh}} \propto (k/k_{\rm rh})^2$ when $k \lesssim k_{\rm rh}$, so the amplitude of the SIGW spectrum is suppressed by a factor of $(k_{\rm rh}/k_*)^2$, which can be easily understood through the following reasoning. Prior to the scalar mode with wavenumber $k_*$ entering the horizon, a tensor mode with wavenumber $k$ experiences a constant source, resulting in a growth proportional to $(k\tau)^2$, where $\tau$ represents the conformal time. However, once the scalar mode $k_*$ reenters the horizon, occurring approximately at $\tau_*\sim1/k_*$, the source effectively ceases, leading to a termination of the growth in the tensor mode. This means that if the amplitude of the SIGW spectrum at the reheating wavenumber remains constant, then as the peak scale moves further away from the reheating scale, the amplitude of the scalar perturbation decreases. An illustration of the energy density spectrum of SIGWs as a function of $w$ can be found in \Fig{ogw}.

\begin{table*}
    \centering
	\begin{tabular}{c|cccccc}
		\hline\hline
		Parameter & $\log_{10} (f_*/\mathrm{Hz})$ & $\log_{10} \Delta$ & $\log_{10} A$ & $\log_{10} (T_{\mathrm{rh}}/\mathrm{Gev})$ & $w$\\[1pt]
		\hline
		 Prior& $\mU(-10, -2)$ & $\mU(-6, -1)$ & $\mU(-5, 1)$ & $\mU(\log_{10} 0.004, 0.6)$ &  $\mU(-1/3, 1.2)$  & \\[1pt]
		Result  & $-5.86^{+2.26}_{-1.03}$ & $\lesssim -2.32$ & $-2.01^{+1.75}_{-1.14}$ & $\lesssim -0.7$ & $0.60^{+0.32}_{-0.39}$\\[1pt]
  \hline
	\end{tabular}
	\caption{\label{tab:priors}Prior distributions and results for the model parameters. Here $\mU$ denotes the uniform distribution. We quote each parameter's median value and $90\%$ equal-tail credible interval.}
\end{table*}

PBHs are formed through gravitational collapse when the density contrast $\delta\rho/\rho$ surpasses a threshold $\delta_c$ in Hubble patches depending on the EoS parameter $w$ at the moment of re-entry~\cite{Zeldovich:1967lct,Hawking:1971ei,Carr:1974nx,Meszaros:1974tb,Carr:1975qj,Musco:2004ak,Musco:2008hv,Musco:2012au,Harada:2013epa,Escriva:2020tak}. For the purpose of a cautious estimation, we adopt Carr's criterion in the uniform Hubble slice~\cite{Carr:1975qj}, which can be translated to the comoving slice as follows~\cite{Domenech:2020ers,Balaji:2023ehk}
\begin{align}\label{dcb}
\delta_c \simeq \frac{3(1+w)}{5+3w} c_s^2=\frac{3(1+w)}{5+3w}.
\end{align}
{The justification for our choice of the critical density threshold $\delta_c$ lies in the fact that the fluctuations of the scalar field propagate with a speed of sound squared. This property makes the formation of PBHs more challenging compared to an adiabatic perfect fluid. Hence, we adopt the upper limit of the density threshold as suggested by Ref.~\cite{Harada:2013epa} \footnote{{It is worth mentioning that the case of general $c_s^2=w$ has been investigated in Ref.~\cite{Musco:2012au}, but it is not feasible to extrapolate their analysis to the specific case of $c_s^2=1$.}}. }

The abundance of PBHs at the time of their formation can be determined using the Press-Schechter theory~\cite{Sasaki:2018dmp} as
\begin{equation}
    \beta(M)=\frac{\gamma}{2}\text{erfc}\left(\frac{\delta_c(w)}{\sqrt2\sigma(M)}\right),
\end{equation}
where $\gamma\approx0.2$ represents the fraction of matter within the Hubble horizon that undergoes collapse to form PBHs. The quantity $\sigma(M)$ denotes the variance of the density perturbation smoothed over the mass scale of $M$. Using the relation between curvature perturbation $\mathcal{R}$ and density contrast $\delta$, 
\begin{equation}
    \delta(t, q)=\frac{2(1+w)}{5+3 w}\left(\frac{q}{a H}\right)^2 \mathcal{R}(q), 
\end{equation}
we can compute $\sigma(M)$ as 
\begin{equation}
\label{sigma}
\sigma^{2} =\left(\frac{2+2w}{5+3w}\right)^{\!\!2}\!\!\! \int_{0}^{\infty}\! \frac{\mathrm{d}q}{q}\, \tilde{W}^{2}(\frac{q}{k})\left(\frac{q}{k}\right)^{4} T^{2}(\frac{q}{k}) \mathcal{P}_{\mathcal{R}}(q),
\end{equation}
where $\tilde{W}(q/k)=\exp(-q^{2}/k^{2}/2)$ is the Gaussian window function and $T(q/k)=3(\sin l - l \cos l)/l^{3}$ is the transfer function with $l = q/k/\sqrt{3}$. 

In a general $w$-dominated Universe, the masses of PBHs are connected to the comoving scale by
\begin{align}\label{mpbh}
\frac{M_{\rm PBH}}{M_\odot}&\approx  0.01\frac{\gamma}{0.2}\!\left(\frac{k_{\rm rh}}{k_*}\right)^{\!\!\!\frac{3(1+w)}{1+3w}}\!\!\! \left(\frac{106.75}{g_{*r}(T_{\rm rh})}\right)^{\!\!\frac12}\!\!\!\left(\frac{{\rm GeV}}{T_{\rm rh}}\right)^{\!\!\! 2}.
\end{align}
In the case of a very sharp peak in the primordial scalar power spectrum, it results in a monochromatic mass function for the PBHs. The abundance of PBHs expressed as the PBH energy fraction with respect to cold dark matter is described by
\begin{equation}\label{fpbh}
f_{\rm PBH}\equiv\frac{\Omega_\text{PBH}}{\Omega_\text{CDM}} \approx 1.5\times 10^{13}\beta\left(\frac{k_*}{k_{\rm rh}}\right)^{\frac{6w}{1+3w}}\left(\frac{T_{\rm rh}}{{\rm GeV}}\right) \left(\frac{g_{*s}(T_{\rm rh})}{106.75}\right)^{-1}\left(\frac{g_{*r}(T_{\rm rh})}{106.75}\right).  
\end{equation}
Each mode with wavenumber $k$ crosses the horizon at the temperature $T$ can be estimated as
\begin{equation}
\label{k-T}
    k\! \simeq \frac{\!1.5\!\times\!  10^7}{\mathrm{Mpc}}
    \left({g_{*r}(T)} \over {106.75} \right)^{1\over 2}\!\!
    \left({g_{*s}(T)} \over 106.75\right)^{-\frac{1}{3}}\!\!
    \left (\frac{T}{\rm GeV}\right).
\end{equation}
The corresponding GW frequency $f$ at current for each mode $k$ is given by 
\begin{equation}
\label{k-f}
    f= \frac{k}{2 \pi} \simeq 1.6\, {\rm nHz}
    \left( \frac{k}{10^6\,{\rm Mpc}^{-1}} \right).
\end{equation}
By combining Eqs. \eqref{k-T} and \eqref{k-f}, we obtain the relation between the frequency $f$ and the temperature $T$ as
\begin{equation}
    f \simeq 24\, {\rm nHz} \left({g_{*r}(T)} \over {106.75} \right)^{1\over 2}\!\!
    \left({g_{*s}(T)} \over 106.75\right)^{-\frac{1}{3}}\!\!
    \left (\frac{T}{\rm GeV}\right).
\end{equation}
Notice that the lower bound of reheating  temperature for BBN is $T_{\mathrm{rh}}\geq 4\,{\rm MeV}$~\cite{Kawasaki:1999na,Kawasaki:2000en,Hannestad:2004px,Hasegawa:2019jsa}, hence the constraints of reheating frequency $f_{\mathrm{rh}}$ for the reheating mode $k_{\mathrm{rh}}$ is $f_{\mathrm{rh}}\gtrsim 0.1 \, {\rm nHz}$, which is smaller than the sensitive frequencies of PTAs.

\section{Data analyses and results}

In this work, we jointly use the NANOGrav 15-year data set, PPTA DR3, and EPTA DR2 to estimate the model parameters. Specifically, we use the free spectrum amplitudes derived by each PTA when considering the Hellings-Downs spatial correlations. Given the time span $T_{\mathrm{obs}}$ of a PTA, the frequencies to which the PTA is sensitive start from $1/T_{\mathrm{obs}}$. Specifically, NANOGrav, PPTA and EPTA use $14$~\cite{NANOGrav:2023gor}, $28$~\cite{Reardon:2023gzh}, and $24$~\cite{Antoniadis:2023rey} frequencies when searching for the SGWB signal, respectively. Combining these PTA data yields $66$ frequency components of a free spectrum starting from $1.28$~nHz to $49.1$~nHz. In \Fig{ogw}, we show the data used in the analyses and the energy density of SIGWs by taking different values of $w$. 

We use the posterior data of time delay $d(f)$ released by each PTA that is related to the power spectrum $S(f)$ through
\begin{equation}
d(f)=\sqrt{S(f) / T_{\mathrm{obs}}}.
\end{equation}
Then we get the free spectrum energy density by
\begin{equation}
\hat{\Omega}_{\mathrm{GW}}(f)=\frac{2 \pi^2}{3 H_0^2} f^2 h_c^2(f) = \frac{8\pi^4}{H_0^2} T_{\mathrm{obs}} f^5 d^2(f),
\end{equation}
where the characteristic strain, $h_c(f)$ is defined by
\begin{equation}
h_c^2(f)=12 \pi^2 f^3 S(f).
\end{equation}
For each observed frequency $f_i$, with the posteriors of $\hat{\Omega}_{\mathrm{GW}}(f_i)$ given above, we can estimate the corresponding kernel density $\mathcal{L}_i$. Therefore, the total log-likelihood is 
\e 
\ln \mathcal{L}(\Lambda) = \sum_{i=1}^{66} \ln \mathcal{L}_i(\Omega_{\mathrm{GW}}(f_i, \Lambda)),
\q
where $\Lambda\equiv \{A, \Delta, f_*, T_{\mathrm{rh}}, w\}$ denotes the set of five model parameters.
We utilize the \texttt{dynesty}~\cite{Speagle:2019ivv} sampler available in the \texttt{Bilby}~\cite{Ashton:2018jfp,Romero-Shaw:2020owr} package to explore the parameter space. We summarize the priors and results for the model parameters in \Table{tab:priors}.

\begin{figure}[tbp!]
	\centering
 \includegraphics[width=\textwidth]{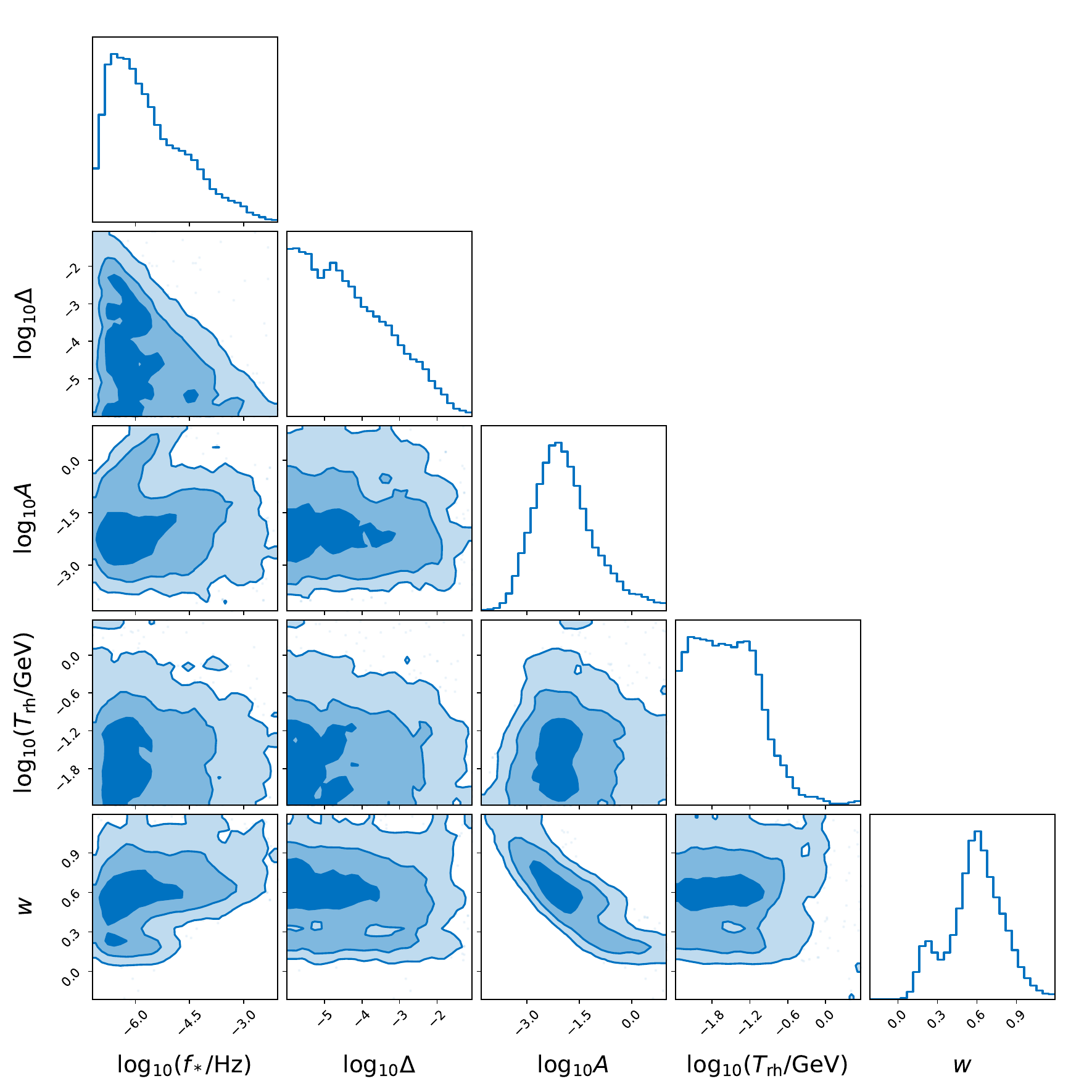}
	\caption{\label{posts_w}One and two-dimensional marginalized posteriors for the model parameters obtained from the combined NANOGrav 15-year data set, PPTA DR3, and EPTA DR2. The contours in the two-dimensional plot correspond to the $1 \sigma$, $2 \sigma$, and $3 \sigma$ credible regions, respectively.}
\end{figure}

\begin{figure}[tbp!]
	\centering
 \includegraphics[width=\textwidth]{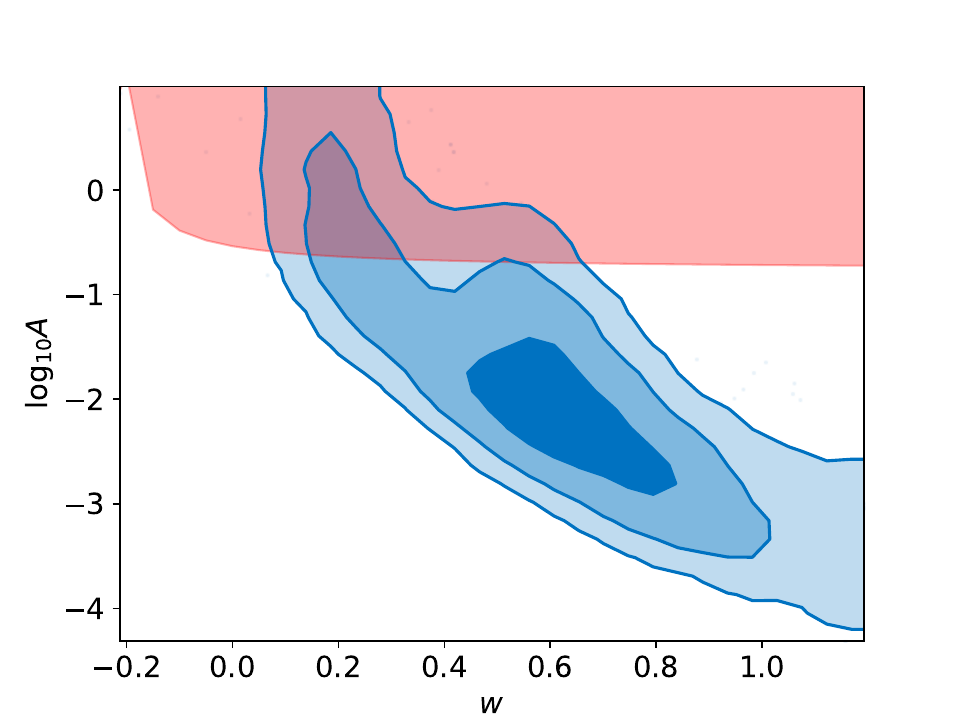}
	\caption{\label{post_with_fpbh} The two dimensional posteriors for the $w$ and $A$ parameters shown in blue contours and the parameter space where $\fpbh > 1$ shown in red region. Other model parameters are fixed as $\log_{10} (f_*/\mathrm{Hz}) = -5.86$, $\log_{10} \Delta=-4.54$, $\log_{10} (T_{\mathrm{rh}}/\mathrm{Gev}) = -1.6$.}
\end{figure}

We show the posteriors for the model parameters in \Fig{posts_w}. We find that the PTA data prefer a narrow curvature power spectrum with $\Delta \lesssim 0.005$ at $95\%$ confidence level when considering the effect of EoS, thus validating our assumption of a narrow peak. Meanwhile, the peak frequency is constrained to be $\log_{10} (f_*/\mathrm{Hz})=-5.86^{+2.26}_{-1.03}$. Furthermore, the reheating temperature has an upper bound that $T_\mathrm{rh} \lesssim 0.2\,\mathrm{GeV}$. Note that we require $T_\mathrm{rh} \gtrsim 4\,\mathrm{MeV}$ to meet the BBN's constraint. Moreover, the amplitude is $\log_{10} A = -2.01^{+1.75}_{-1.14}$, and the EoS parameter is $w = 0.60^{+0.32}_{-0.39}$. Although we allow the prior $w$ to span the range of $[-1/3, 1.2]$, the result excludes the negative EoS parameter at $95\%$ confidence level and prefers a value of EoS parameter that $w < 1$. Nevertheless, $w=1/3$ is consistent with the PTA data. An inspection of the 1-D marginalized posterior of $w$ shows a minor peak in $w\sim 0.19$ in addition to the major peak $w\sim 0.58$ mainly because small and larger EoS parameters are degenerate. As argued in Refs.~\cite{Domenech:2020kqm,Domenech:2021ztg}, the low-frequency tail of SIGWs generated from a narrow peaked primordial spectrum is $f^{2-2|b|}$. For $w \approx 0.19$ and $w \approx 0.58 $, we both have $2-2|b|\approx 1.73$. The amplitude $A$ helps to break this degeneracy partially. It is worth mentioning that the relatively larger value of the EoS parameter preferred by PTA data will help to avoid the PBH overproduction, as can be seen from \Fig{post_with_fpbh}, where the red region denotes the excluded parameter space by the requirement that $\fpbh \leq 1$. {The negative correlations observed between $w$ and $A$ in Fig.~3 can be explained as follows. Referring to \Eq{eq:OmegaDelta}, with fixed values of $k$, $k_*$ and $\Delta$, we have $\Omega_{\rm GW,0}h^2 \propto \Omega^\delta_{\rm GW,0}h^2$. By further evaluating \Eq{ogwrh}, we have $\Omega^\delta_{\rm GW,0}h^2 \propto A^2\, (k/k_\mathrm{rh})^{-2b}\, \frac{\mathcal{T}(\frac{k_*}{k}, \frac{k_*}{k}, w)}{(k/k_*)^2} \Theta(2-\frac{k}{k_*})$, where $\Theta$ is Heaviside step function. For the case where $k_\mathrm{rh}\ll k \ll k_*$ and $w>0$, it becomes evident that an increase in $w$ results in a corresponding increase in $\Omega^\delta_{\rm GW,0}h^2$, thus necessitating a smaller value of $A$ to compensate for it.}

\section{Summary and discussion}

We measure the EoS parameter of the early Universe through SIGWs inspired by the recent evidence of the nanohertz stochastic signal reported by the PTA experiments. 
With a comprehensive analysis of the data from NANOGrav, PPTA, and EPTA, we find the preferred value of $w$ peaks at around $0.6$, implying that SIGWs are produced during the preheating process where the homogeneous inflaton oscillates or rolls down the effective potential. Since during the oscillation of inflaton, $w=\frac{p-2}{p+2}$ for an power-law potential $V(\phi)\propto \phi^{p}$~\cite{Johnson:2008se,Turner:1983he}, then, the constraint on $w$ implies a $\phi^{8}$ bottom of the inflationary potential. 
The PTA data also implies that $T_\mathrm{rh}$ should not exceed $0.2$~GeV, safely allowing a successful BBN.

Our result is also applicable to other cases of condensate domination, such as the early dark energy and the oscillation of spectators after inflation, which are proposed to alleviate the Hubble tension~\cite{Poulin:2018cxd,Niedermann:2020dwg} and generate primordial curvature perturbations~\cite{Lyth:2001nq}, respectively. Since the energy density of the condensate with $w\sim0.6$ decreases more quickly than radiation, the Universe soon returns to the radiation-dominated era, and the subsequent evolution remains unchanged. {Nevertheless, we should emphasize that the radiation domination ($w=1/3$) is consistent with the PTA data within the $90\%$ credible interval.}

\section*{Note added}
While completing this work, we found a parallel independent work~\cite{Zhao:2023joc} where the authors utilized the NANOGrav 15-year data set in conjunction with CMB and BBN and found a weak constraint on the EoS parameter that has no upper bound. As a comparison, we incorporated all available data from NANOGrav, PPTA, and EPTA and constrained the EoS parameter to be $w = 0.60^{+0.32}_{-0.39}$.
	
\section*{Acknowledgments}
We are very grateful to Jing Liu for the fruitful discussions that greatly improve the manuscript.
LL is supported by the National Natural Science Foundation of China (Grant No.~12247112 and No.~12247176) and the China Postdoctoral Science Foundation Fellowship No.~2023M730300. 
ZCC is supported by the National Natural Science Foundation of China (Grant No.~12247176 and No.~12247112) and the China Postdoctoral Science Foundation Fellowship No.~2022M710429. QGH is supported by grants from NSFC (Grant No.~12250010, 11975019, 11991052, 12047503), Key Research Program of Frontier Sciences, CAS, Grant No.~ZDBS-LY-7009, CAS Project for Young Scientists in Basic Research YSBR-006, the Key Research Program of the Chinese Academy of Sciences (Grant No.~XDPB15). 

\bibliographystyle{JHEP}
\bibliography{refs}

\end{document}